\documentclass[sigconf]{acmart}

\usepackage{tikz}
\usepackage{fontawesome5}
\usetikzlibrary{patterns}
\usetikzlibrary{arrows.meta, fit, positioning}

\AtBeginDocument{%
  }

\copyrightyear{2026}
\acmYear{2026}
\setcopyright{cc}
\setcctype{by}
\acmConference[CIKM '26]{Proceedings of the 35th ACM International Conference on Information and Knowledge Management}{November 07--11, 2026}{Rome, Italy}
\acmBooktitle{Proceedings of the 35th ACM International Conference on Information and Knowledge Management (CIKM '26), November 07--11, 2026, Rome, Italy}
\acmISBN{979-8-4007-2539-5/2026/11}
\acmDOI{10.1145/3799682.3840209}

\begin{document}

%%
%% The "title" command has an optional parameter,
%% allowing the author to define a "short title" to be used in page headers.
\title[NewsReX: An Open-Source Multi-Framework for Neural News Recommendation]{{\itshape NewsReX}: An Open-Source Multi-Framework for Neural News Recommendation}

\author{Igor L.R. Azevedo}
\orcid{0000-0001-5144-825X}
\affiliation{%
  \institution{Imperial College London}
  \city{London}
  \country{United Kingdom}
}
\email{igorazevedo@acm.org}

\author{Toyotaro Suzumura}
\orcid{0000-0001-6412-8386}
\affiliation{%
  \institution{The University of Tokyo}
  \city{Tokyo}
  \country{Japan}
}
\email{suzumura@acm.org}

\author{Yuichiro Yasui}
\orcid{0000-0002-4175-9318}
\affiliation{%
  \institution{Nikkei Inc.}
  \city{Tokyo}
  \country{Japan}
}
\email{yuichiro.yasui@nex.nikkei.com}

\begin{abstract}
    Reproducing and comparing results in neural news recommendation research remains difficult due to fragmented codebases, inconsistent preprocessing pipelines, and tight coupling between models and deep learning frameworks. We present NewsReX\footnote{\url{https://github.com/igor17400/NewsReX}}, an open-source framework that unifies multiple neural news recommendation models across two ML backends (PyTorch and JAX/Flax). Models are defined through a structured YAML specification that captures architecture, training recipe, and data contract, decoupling model definition from framework-specific implementation. A Hydra-based configuration system enables composable experiments with single-command framework switching. We provide cross-framework parity tests to verify numerical consistency across backends, allowing researchers to reproduce results from the original implementations and then transfer experiments to their preferred framework. We report comprehensive experiments on the MIND-small benchmark with multiple seeds per model and per framework, and demonstrate multilingual generalization by training on a proprietary Japanese news dataset from Nikkei News\footnote{\url{https://nikkei.com/}}. We also publish pre-trained model weights on Hugging Face\footnote{\url{https://huggingface.co/newsrex/}} so that researchers can build on existing results without redundant retraining, reducing both computational cost and energy consumption. All source code, experiment configurations, and trained model weights are publicly available.
\end{abstract}




\maketitle

\section{Introduction}

Neural news recommendation (NNR) has become the dominant paradigm for personalized news delivery, with deep learning models consistently outperforming traditional collaborative filtering and content-based approaches. Yet reproducing and fairly comparing NNR results remains a persistent challenge: reimplementing models is time-consuming, preprocessing pipelines vary across codebases, and critical details such as hyperparameter settings and evaluation protocols are frequently omitted or inconsistent.

\begin{figure*}[t]
    \centering
    \begin{tikzpicture}[scale=0.88, every node/.style={transform shape},
        box/.style={draw, rounded corners=4pt, minimum height=0.85cm, align=center, font=\small},
        spec/.style={box, fill=blue!8, minimum width=2.8cm, font=\small\bfseries},
        hydra/.style={box, fill=orange!10, minimum width=2.8cm},
        core/.style={box, fill=green!8, minimum width=2.8cm},
        fw/.style={box, fill=red!8, minimum width=2.2cm, minimum height=0.75cm, font=\small},
        adapter/.style={box, dashed, fill=yellow!8, minimum width=2.0cm, font=\small\itshape},
        eval/.style={box, fill=green!8, minimum width=2.8cm},
        hf/.style={box, fill=yellow!15, minimum width=2.8cm},
        use/.style={box, fill=gray!8, minimum width=2.0cm, font=\scriptsize},
        groupbox/.style={draw, rounded corners=6pt, dashed, thick, gray},
        arrow/.style={-{Stealth[length=2.5mm]}, thick},
        lbl/.style={font=\scriptsize\bfseries\color{gray!70!black}, align=center},
    ]

    % === Logo (lower left, in the empty band beside the downstream row) ===
    \node[anchor=center] (logo) at (4, -2.2) {\includegraphics[width=3.2cm]{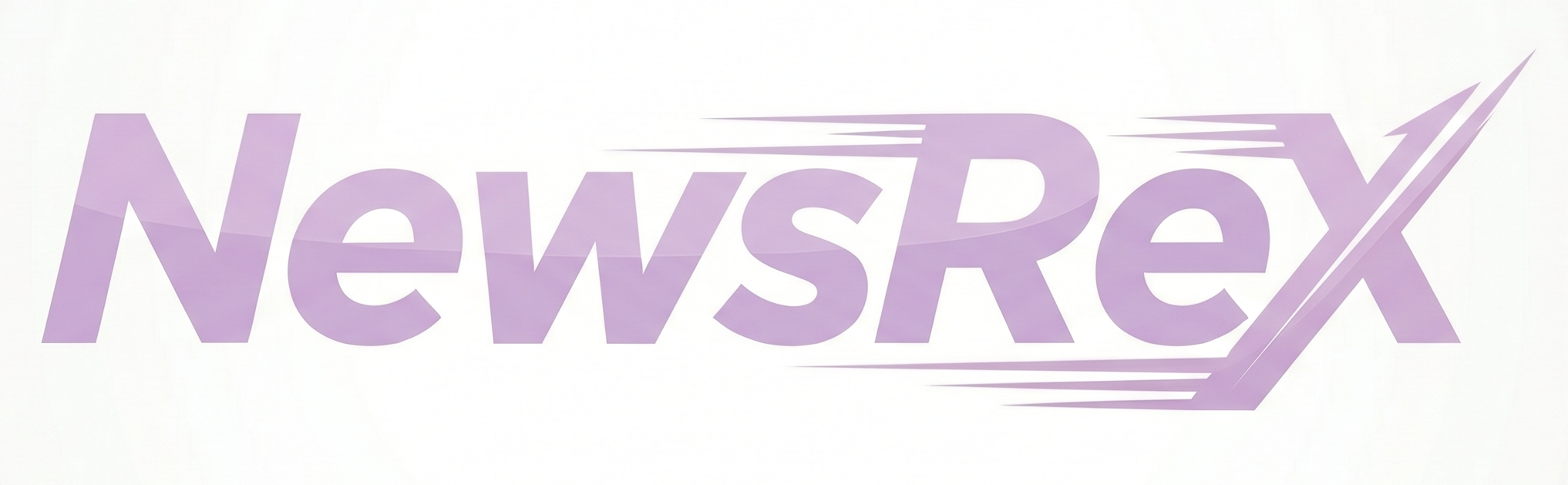}};

    % === Spec + Hydra stacked ===
    \node[spec] (spec) at (3.8, 1.4) {\faFileCode\; Spec (YAML)\\[-1pt]{\tiny architecture $\cdot$ recipe $\cdot$ data}};
    \node[hydra] (hydra) at (3.8, 0.1) {\faCogs\; Hydra Config\\[-1pt]{\tiny base + spec + experiment}};

    % Dashed box around both
    \node[groupbox, fit=(spec)(hydra), inner sep=5pt, label={[lbl, anchor=south]above:{\tiny \texttt{framework=jax}
    $\vert$ \texttt{pytorch}}}] (configgroup) {};

    % Arrow between spec and hydra
    \draw[arrow] (spec) -- (hydra);

    % === Data Processing ===
    \node[core] (data) at (7.5, 0.8) {\faDatabase\; Data\\Processing\\[-1pt]{\tiny tokenize $\cdot$ embed $\cdot$ sample}};

    % === Frameworks ===
    \node[fw] (pytorch) at (10.8, 1.4) {\faFire\; PyTorch {\tiny \texttt{nn.Module}}};
    \node[fw] (jax)     at (10.8, 0.2) {\faBolt\; JAX/Flax {\tiny \texttt{Module}}};

    \node[groupbox, fit=(pytorch)(jax), inner sep=5pt] (fwgroup) {};

    % === Adapter ===
    \node[adapter] (adapter) at (13.8, 0.8) {\faRandom\; Adapter\\[-1pt]{\tiny $\rightarrow$ NumPy}};

    % === Evaluation ===
    \node[eval] (eval) at (17.0, 0.8) {\faChartBar\; Evaluation\\[-1pt]{\tiny AUC $\cdot$ MRR $\cdot$ NDCG@k}};

    % === HuggingFace Hub (below, connected from eval) ===
    \node[hf] (hf) at (17.0, -1.0) {\faCloud\; HuggingFace Hub\\[-1pt]{\tiny weights $\cdot$ config $\cdot$ model card}};

    % === Downstream uses ===
    \node[use] (infer) at (14.5, -2.5) {\faSearch\; Inference\\[-1pt]{\tiny no GPU needed}};
    \node[use] (finetune) at (17.0, -2.5) {\faCog\; Fine-tune\\[-1pt]{\tiny new dataset}};
    \node[use] (bench) at (19.5, -2.5) {\faChartLine\; Benchmark\\[-1pt]{\tiny compare models}};

    % === Arrows (main flow) ===
    \draw[arrow] (configgroup.east |- data.west) -- (data.west);
    \draw[arrow] (data) -- (pytorch.west);
    \draw[arrow] (data) -- (jax.west);
    \draw[arrow] (pytorch.east) -- (adapter.north west);
    \draw[arrow] (jax.east) -- (adapter.south west);
    \draw[arrow] (adapter) -- (eval);

    % === Arrow: eval to HF ===
    \draw[arrow] (eval) -- (hf) node[midway, right, font=\tiny, text=gray] {publish};

    % === Arrows: HF to downstream ===
    \draw[arrow] (hf) -- (infer);
    \draw[arrow] (hf) -- (finetune);
    \draw[arrow] (hf) -- (bench);

    % === Labels below ===
    \node[lbl, below=4pt] at (configgroup.south) {Configuration};
    \node[lbl, below=4pt] at (fwgroup.south) {Training\\(framework-specific)};
    \node[lbl, above=4pt] at (eval.north) {Evaluation\\(framework-agnostic)};

    \end{tikzpicture}
    \caption{\small Overview of the \textit{NewsReX} architecture. Models are defined via YAML specs, composed through Hydra, and trained in either PyTorch or JAX/Flax. A lightweight adapter converts representations to NumPy for framework-agnostic evaluation.}
    \Description{Block diagram of the NewsReX pipeline. A YAML model specification feeds a Hydra configuration, which selects either the PyTorch or the JAX/Flax backend. Shared data processing supplies both backends; each backend's trained encoders pass through a lightweight adapter that converts outputs to NumPy for a single framework-agnostic evaluation stage computing AUC, MRR and NDCG@k. Trained weights are published to the Hugging Face Hub, from which they can be used for inference, fine-tuning, or benchmarking.}
    \label{fig:architecture}
  \end{figure*}

The earliest neural architectures for news recommendation such as NRMS~\cite{nrms}, NAML~\cite{naml}, and LSTUR~\cite{lstur} were released as part of the \textbf{Microsoft Recommenders} library~\cite{recommenders-microsoft} in Keras TensorFlow~\cite{tensorflow2015-whitepaper}. These models remain heavily cited as baselines across the literature. However, running them today using this framework is hindered by library version incompatibilities and degraded performance relative to modern tooling. Subsequent models such as PP-Rec~\cite{pprec}, CAUM~\cite{caum}, and TCCM~\cite{tccm} were published at top venues yet see comparatively little adoption as baselines, all of them implemented in Keras \cite{chollet2015keras}. More recent models such as GLORY~\cite{glory}, CROWN~\cite{crown}, and DIGAT~\cite{digat} were implemented in PyTorch \cite{pytorch}. This fragmentation carries a deeper cost for open reproduction. When code is unavailable or difficult to run, errors in model implementations or evaluation can go undetected. \textbf{NewsRecLib}~\cite{newsreclib}, built on PyTorch Lightning~\cite{Falcon_PyTorch_Lightning_2019} and Hydra~\cite{Yadan2019Hydra}, has made significant progress toward solving this problem by offering a unified, configuration-driven framework for reproducible evaluation in PyTorch. More general configuration-driven libraries also cover news recommenders: \textbf{RecBole}~\cite{recbole} across recommendation families, and \textbf{Legommenders}~\cite{legommenders} for content-based models with LLM support. Neither exposes a non-PyTorch backend, and both treat NNR as one case of a broader taxonomy rather than modelling its splitting, negative sampling, and impression-level evaluation protocols directly.

\textit{\textbf{NewsReX}} builds on this direction by extending multi-model support to two backends---PyTorch and JAX/Flax~\cite{jax2018github}---behind a unified interface, leveraging the complementary strengths of each framework. JAX's JIT compilation delivers significant speedups for attention- and convolution-based models (e.g., NRMS, NAML, LSTUR), while PyTorch remains faster for GNN-based architectures (e.g., GLORY, CROWN, DIGAT) thanks to its mature graph ecosystem, including PyTorch Geometric~\cite{pygeo}. 

To our knowledge, \textit{NewsReX} is the first library to provide JAX implementations of NNR models and to offer shared, reusable graph construction infrastructure that simplifies adding new GNN-based recommendation models. Its key contributions are:

\begin{enumerate}
  \item \textbf{Framework-Agnostic Model Specification.} A declarative YAML spec captures architecture, training recipe, and data contract, fully decoupling model logic from framework-specific code.
  \item \textbf{Cross-Framework Parity.} After training, encoders export representations to NumPy via a lightweight adapter; all scoring and metrics are computed in shared, backend-independent code.
  \item \textbf{Pre-Trained Weights.}\ Trained model weights are published on Hugging Face, enabling reproduction and extension of results through inference alone.
  \item \textbf{Open by Default.} All code, configurations, and artifacts are MIT-licensed, ensuring every implementation decision is inspectable and auditable.
\end{enumerate}

In addition to these distinguishing features, \textit{NewsReX} incorporates established best practices: Hydra-based configurability for full experiment reproducibility \cite{Yadan2019Hydra}, multilingual embedding support, integration with Weights~\&~Biases~\cite{wandb} for experiment tracking, and a modular architecture that simplifies the addition of new models and components. All source code and experiment configurations are publicly available at \url{https://github.com/igor17400/NewsReX} and model weights at \url{https://huggingface.co/newsrex/}.

\section{\textit{NewsReX} Framework}

\textit{NewsReX} separates \emph{what} a model computes from \emph{how} it is executed. Each model is defined by a
YAML specification that declares its architecture, training recipe, and data contract. Framework-specific runners
(PyTorch, JAX/Flax) consume this specification and instantiate the corresponding implementation, ensuring
that researchers interact with a single model definition regardless of backend. Figure~\ref{fig:architecture} illustrates this pipeline.

\subsection{Model Specification}

A model specification captures three concerns in a declarative format: \textbf{(1) Architecture}: encoder types (e.g., multi-head self-attention, CNN, GRU), dimensions, number of attention heads, dropout rates, and click predictor type. \textbf{(2) Training recipe}: optimizer, learning rate, loss function, negative sampling ratio, and early stopping criteria. \textbf{(3) Data contract}: input fields (title, abstract, category), maximum sequence lengths, and history size. This declarative layer decouples model logic from framework-specific code. Adding a new model requires writing a spec and the corresponding encoder implementations in each target framework. Shared data processing, evaluation, and configuration infrastructure are reused automatically.

\subsection{Encoders}

Following the standard methodology established by foundational NNR models~\cite{nrms, naml, lstur}, every
recommendation model in \textit{NewsReX} is composed of at least two primary components: a News Encoder and a User Encoder.

\subsubsection{News Encoder}

The News Encoder produces a fixed-size embedding $\mathbf{e}_n$ for a given news article. Let a news title be
represented as a sequence of word embeddings $\mathbf{E}_T = [\mathbf{e}_{w_1}, \dots, \mathbf{e}_{w_L}]$. A text
encoder (e.g., CNN~\cite{naml} or multi-head self-attention~\cite{nrms}) processes this sequence, followed by additive
attention to aggregate word representations:

\begin{align}
s_i &= \mathbf{v}_n^{\top} \tanh(\mathbf{W}_1 \mathbf{e}_{w_i} + \mathbf{b}_1), \quad
a_i = \frac{\exp(s_i)}{\sum_{j=1}^{L} \exp(s_j)}
\end{align}

The final news embedding is $\mathbf{e}_n = \sum_{i=1}^{L} a_i \, \mathbf{e}_{w_i}$. More advanced encoders incorporate additional views such as abstract, category, subcategory~\cite{naml, caum} or external knowledge such as entity embeddings~\cite{digat}, global popularity signals~\cite{pprec, tccm}, and pre-trained language model representations (e.g., BERT~\cite{devlin2019bert}) for richer contextual encoding.

\subsubsection{User Encoder}

The User Encoder produces a user representation $\mathbf{u}$ by aggregating embeddings of previously clicked articles $\{\mathbf{e}_{n_1}, \dots, \mathbf{e}_{n_K}\}$. Common strategies include multi-head self-attention~\cite{nrms}, GRU-based sequential modeling~\cite{lstur}, candidate-aware attention~\cite{caum}, and graph neural networks that capture inter-article relationships~\cite{glory, crown}.

\subsubsection{Click Prediction}

The relevance score between a user $\mathbf{u}$ and a candidate article $\mathbf{e}_c$ is computed via dot product:
$\hat{y} = \mathbf{u}^{\top} \mathbf{e}_c$. During training, scores over one positive and $N$ negative samples are optimized with cross-entropy loss:
\begin{equation}
\mathcal{L} = -\log \frac{\exp(\hat{y}^+)}{\exp(\hat{y}^+) + \sum_{i=1}^{N} \exp(\hat{y}_i^-)}
\end{equation}
\textit{NewsReX} decouples the loss from the model definition, so alternative objectives can be implemented without modifying model code.

\subsection{Multi-Framework Execution}

\textit{NewsReX} enforces a strict separation: frameworks are used exclusively for training, while evaluation is
entirely framework-agnostic. The architecture is organized as follows:

\begin{itemize}
    \item \textbf{Framework-specific training} (\texttt{src/frameworks/\{pytorch, jax\}}). Each backend provides a runner that orchestrates its native training loop, model implementations that translate spec parameters into native layers (e.g., \texttt{nn.Module}, Flax \texttt{Module}), and dataloaders adapted to each framework's batching conventions.

    \item \textbf{Adapter layer.} Each framework implements a lightweight adapter that exposes two operations: encode news and encode users. The adapter handles device placement and framework specific inference conventions (e.g., \texttt{torch.no\_grad()}, JAX JIT), then converts all outputs to NumPy arrays.

    \item \textbf{Shared evaluation} (\texttt{src/core/}). Once representations are in NumPy, a single evaluation pipeline scores impressions via dot products and computes all metrics (AUC, MRR, NDCG@k) without any framework dependency. 
\end{itemize}

The design is focused on a framework-agnostic paradigm as much as possible at the specification, configuration, and evaluation levels: model configs, data preprocessing, and all metrics live in shared, NumPy-only code. The only framework-specific code resides in the encoder implementations and their thin adapter wrappers. For non-GNN models (NRMS, NAML, LSTUR, PP-Rec, CAUM, TCCM), the PyTorch and JAX implementations are structurally equivalent translations of the same spec.

For GNN-based models (GLORY, CROWN, DIGAT), \textit{NewsReX} separates graph \emph{construction} from graph \emph{computation}. All graph construction, including co-occurrence graphs, user-topic adjacency matrices, semantic-augmented graphs (SAG), neighbor dictionaries, and subgraph sampling, is implemented once in shared code, so adding a new GNN model to either backend requires only the forward-pass graph layers, not the data pipeline. The graph computation layer (message passing, scatter operations) is necessarily framework-specific. In our experience, although JAX can match or exceed PyTorch's speed for certain GNN models, the implementation cost is considerably higher: PyTorch Geometric~\cite{pygeo} provides optimized, well-tested primitives (e.g., \texttt{GatedGraphConv}, scatter operations) that make prototyping new GNN architectures straightforward, whereas JAX equivalents must be hand-rolled from scratch. For this reason, we recommend PyTorch as the default backend for GNN-based models in \textit{NewsReX}.

Switching frameworks requires changing a single Hydra parameter (\texttt{framework=jax,pytorch}); all other configurations (model spec, dataset, hyperparameters, and evaluation) remain identical.

\subsection{Configuration and Experiment Composition}

Experiments are composed through Hydra's override system at three levels: (1)~a base configuration defining training defaults and device settings, (2)~a model spec defining architecture and recipe, and (3)~an experiment preset wiring a spec to a dataset with optional overrides. This layered approach ensures that every experimental setting is captured in version-controlled YAML files, enabling exact reproduction with a single command.

\section{Experiments} \label{sec:experiments}

We evaluate \textit{NewsReX} on three axes: \textbf{(1)} recommendation accuracy across models and frameworks,
\textbf{(2)} cross-framework consistency, and \textbf{(3)} training efficiency. All models are trained on the
MIND-small dataset~\cite{wu2020mind} using the default spec for each model for up to 10 epochs with early stopping (patience of 3 epochs). Results report the mean and standard deviation over different seeds. Metrics follow standard NNR evaluation: AUC, MRR, NDCG@5, and NDCG@10. For every model card uploaded to Hugging Face, we have detailed the exact parameters used to train the model for transparency.

\begin{table*}[t]
  \centering
  \caption{\small Cross-framework comparison of non-GNN models on MIND-small (GloVe~\cite{pennington-etal-2014-glove}, \texttt{random} split). Mean $\pm$ std over 3 seeds (42, 123, 456).}
  \label{tab:mind_results}
  \small
  \begin{tabular}{l|cccc|cccc}
  \toprule
   & \multicolumn{4}{c|}{\textbf{PyTorch}} & \multicolumn{4}{c}{\textbf{JAX/Flax}} \\
  \textbf{Model} & \textbf{AUC} & \textbf{MRR} & \textbf{nDCG@5} & \textbf{nDCG@10} & \textbf{AUC} & \textbf{MRR} & \textbf{nDCG@5} & \textbf{nDCG@10} \\
  \midrule
  NRMS   & 65.44{\tiny$\pm$.37} & 30.90{\tiny$\pm$.49} & 33.90{\tiny$\pm$.52} & 40.37{\tiny$\pm$.46} & 65.85{\tiny$\pm$.28} & 31.28{\tiny$\pm$.18} & 34.37{\tiny$\pm$.25} & 40.78{\tiny$\pm$.21} \\
  NAML   & 66.67{\tiny$\pm$.33} & 31.97{\tiny$\pm$.09} & 35.29{\tiny$\pm$.23} & 41.59{\tiny$\pm$.18} & 66.03{\tiny$\pm$.56} & 31.13{\tiny$\pm$.39} & 34.37{\tiny$\pm$.44} & 40.70{\tiny$\pm$.42} \\
  LSTUR  & 66.51{\tiny$\pm$.24} & 31.60{\tiny$\pm$.28} & 34.97{\tiny$\pm$.37} & 41.18{\tiny$\pm$.30} & 66.72{\tiny$\pm$.27} & 31.76{\tiny$\pm$.24} & 35.11{\tiny$\pm$.30} & 41.44{\tiny$\pm$.22} \\
  PP-Rec & 66.48{\tiny$\pm$.19} & 31.42{\tiny$\pm$.33} & 34.89{\tiny$\pm$.42} & 41.23{\tiny$\pm$.32} & 67.14{\tiny$\pm$.20} & 32.23{\tiny$\pm$.27} & 35.93{\tiny$\pm$.34} & 42.08{\tiny$\pm$.25} \\
  CAUM   & 67.02{\tiny$\pm$.30} & 31.99{\tiny$\pm$.37} & 35.42{\tiny$\pm$.35} & 41.75{\tiny$\pm$.29} & 67.11{\tiny$\pm$.45} & 31.75{\tiny$\pm$.43} & 35.21{\tiny$\pm$.49} & 41.55{\tiny$\pm$.40} \\
  TCCM   & 66.36{\tiny$\pm$.35} & 30.88{\tiny$\pm$.41} & 34.40{\tiny$\pm$.45} & 40.68{\tiny$\pm$.18} & 67.67{\tiny$\pm$.18} & 32.08{\tiny$\pm$.19} & 35.77{\tiny$\pm$.24} & 42.02{\tiny$\pm$.18} \\
  \bottomrule
  \end{tabular}
\end{table*}

\begin{table}[t]
  \centering
  \caption{\small PyTorch GNN models on MIND-small (GloVe, \texttt{random} split). Mean $\pm$ std over 2 seeds (42, 123).}
  \label{tab:gnn_results}
  \footnotesize
  \begin{tabular}{l|cccc}
  \toprule
  \textbf{Model} & \textbf{AUC} & \textbf{MRR} & \textbf{nDCG@5} & \textbf{nDCG@10} \\
  \midrule
  CROWN  & 67.05{\tiny$\pm$.43} & 31.87{\tiny$\pm$.12} & 35.29{\tiny$\pm$.24} & 41.55{\tiny$\pm$.22} \\
  DIGAT  & 67.28{\tiny$\pm$.16} & 32.22{\tiny$\pm$.05} & 35.71{\tiny$\pm$.07} & 41.91{\tiny$\pm$.10} \\
  GLORY  & 65.35{\tiny$\pm$.31} & 30.86{\tiny$\pm$.25} & 34.03{\tiny$\pm$.36} & 40.48{\tiny$\pm$.35} \\
  \bottomrule
  \end{tabular}
\end{table}

\begin{table}[t]
  \centering
  \caption{\small JAX/Flax with BERT embeddings on MIND-small (\texttt{random} split, single seed).}
  \label{tab:jax_plm}
  \footnotesize
  \begin{tabular}{l|cccc}
  \toprule
  \textbf{Model} & \textbf{AUC} & \textbf{MRR} & \textbf{nDCG@5} & \textbf{nDCG@10} \\
  \midrule
  NRMS   & 66.14 & 31.35 & 34.69 & 40.96 \\
  NAML   & 66.59 & 31.48 & 34.93 & 41.33 \\
  LSTUR  & 67.20 & 32.50 & 35.76 & 42.06 \\
  PP-Rec & 66.81 & 33.02 & 36.92 & 42.72 \\
  CAUM   & 68.25 & 32.85 & 36.31 & 42.62 \\
  TCCM   & 67.73 & 33.28 & 37.06 & 43.05 \\
  \bottomrule
  \end{tabular}
\end{table}

\begin{table}[t]
  \centering
  \caption{\small PyTorch with \texttt{dev\_as\_val} split on MIND-small (GloVe, single seed).}
  \label{tab:split_comparison}
  \footnotesize
  \begin{tabular}{l|cccc}
  \toprule
  \textbf{Model} & \textbf{AUC} & \textbf{MRR} & \textbf{nDCG@5} & \textbf{nDCG@10} \\
  \midrule
  NRMS   & 65.51 & 31.32 & 34.33 & 40.73 \\
  NAML   & 66.73 & 32.02 & 35.43 & 41.67 \\
  LSTUR  & 65.73 & 30.53 & 33.89 & 40.24 \\
  CAUM   & 67.41 & 32.61 & 36.04 & 42.32 \\
  \midrule
  CROWN  & 67.22 & 32.29 & 35.88 & 42.03 \\
  DIGAT  & 67.46 & 32.36 & 35.80 & 41.99 \\
  GLORY  & 66.98 & 32.28 & 35.90 & 42.08 \\
  \bottomrule
  \end{tabular}
\end{table}

\subsection{MIND-small Benchmark}

\paragraph{Dataset Split.}

MIND-small~\cite{wu2020mind} is distributed with two directories: \texttt{train/} and \texttt{dev/}. A persistent source of inconsistency in the literature is how these partitions are used for validation and testing. Some published codebases use the official \texttt{dev/} split as both the validation and test set (which we refer to as \texttt{dev\_as\_val}), reporting metrics directly on the partition used for model selection. Others carve a held-out validation set from the training data, reserving \texttt{dev/} exclusively for testing.  \textit{NewsReX} supports three validation strategies via a configuration parameter (\texttt{validation\_split\_strategy}): \textbf{(1)~random}: 5\% of training behaviors are randomly sampled (seed~42) as validation, with \texttt{dev/} used only for testing; \textbf{(2)~chronological}: the last calendar day of training behaviors serves as validation, preserving temporal ordering; \textbf{(3)~dev\_as\_val}: the full \texttt{dev/} partition is used for both validation and testing, with early stopping disabled to avoid information leakage. For the majority of experiments in this paper, we follow the \texttt{random} approach, with the exception of Table~\ref{tab:split_comparison} which reports \texttt{dev\_as\_val} results for comparison.

\paragraph{Negative Sampling.}

During training, each positive click is paired with $K{-}1$ negatives from the same impression (default $K{=}5$). The negative sampling strategy is also configurable: \textit{NewsReX} supports random sampling (default), popularity-weighted sampling, and topic-diverse sampling that distributes negatives across news categories. This allows researchers to study the effect of negative sampling on model performance within a controlled experimental setting. During evaluation, all impression candidates are shuffled to prevent positional bias.

\paragraph{Results.}

Tables~\ref{tab:mind_results}--\ref{tab:split_comparison} illustrate the flexibility of \textit{NewsReX} across frameworks, model families, embedding types, and evaluation protocols. Table~\ref{tab:mind_results} reports non-GNN models trained in both PyTorch and JAX (GloVe, \texttt{random} split, mean $\pm$ std over 3 seeds), confirming cross-framework parity with differences typically within one standard deviation. Table~\ref{tab:gnn_results} demonstrates GNN support via PyTorch, leveraging PyTorch Geometric's graph primitives. Table~\ref{tab:jax_plm} showcases \textit{NewsReX}'s support for pre-trained language model embeddings by reporting JAX results with BERT~\cite{devlin2019bert}. Finally, Table~\ref{tab:split_comparison} provides results under the \texttt{dev\_as\_val} split, since several published works adopt this evaluation protocol, enabling direct comparison.

\subsection{Multilingual Generalization}

To demonstrate that \textit{NewsReX} generalizes beyond English, we train three models (NRMS, NAML, and LSTUR) on a proprietary Japanese news dataset from Nikkei News using BPEmb embeddings~\cite{heinzerling2018bpemb}. Table~\ref{tab:nikkei_results} reports results on this dataset.

\begin{table}[t]
  \centering
  \caption{\small Performance on the proprietary Nikkei News (Japanese) dataset. JAX/Flax, mean $\pm$ std over 3 seeds.}
  \label{tab:nikkei_results}
  \footnotesize
  \begin{tabular}{l|cccc}
  \toprule
  \textbf{Model} & \textbf{AUC} & \textbf{MRR} & \textbf{nDCG@5} & \textbf{nDCG@10} \\
  \midrule
  NRMS   & 60.18{\tiny$\pm$1.20} & 23.96{\tiny$\pm$.55} & 30.02{\tiny$\pm$.79} & 39.84{\tiny$\pm$1.03} \\
  NAML   & 57.81{\tiny$\pm$2.21} & 21.51{\tiny$\pm$1.89} & 27.01{\tiny$\pm$2.81} & 36.62{\tiny$\pm$2.77} \\
  LSTUR  & 59.04{\tiny$\pm$.47} & 21.83{\tiny$\pm$.49} & 27.64{\tiny$\pm$.68} & 37.67{\tiny$\pm$.67} \\
  \bottomrule
  \end{tabular}
\end{table}

\subsection{Training Efficiency}

Figure~\ref{fig:training_time} compares total training time (10 epochs) for PyTorch and JAX across all six non-GNN models on MIND-small (seed 123). JAX is faster overall for every model, with total-time speedups ranging from 1.1$\times$ (LSTUR, CAUM) to 2.2$\times$ (TCCM, NRMS). The per-epoch steady-state speedups are larger---up to 2.8$\times$ for TCCM (130\,s vs.\ 359\,s) and 2.6$\times$ for NRMS (66\,s vs.\ 170\,s)---but JAX pays a one-time JIT compilation cost (hatched region). On this seed the JIT cost is modest for every model, ranging from ${\sim}0.6$\,min (NRMS) to ${\sim}2.4$\,min (TCCM), a small fraction of total time. The smaller total speedups for CAUM and LSTUR stem from JAX's higher per-epoch \emph{evaluation} cost (XLA recompilation over variable impression shapes), which partly offsets its faster training step. These results validate the dual-backend design: researchers can leverage JAX's compilation speedups for attention-based models while using PyTorch's mature ecosystem for GNN architectures.

\begin{figure}[t]
  \centering
  \resizebox{0.98\columnwidth}{!}{%
  \begin{tikzpicture}[xscale=0.155, yscale=1.1]

  % --- Alternating row bands ---
  \fill[gray!5] (-8, 6.6) rectangle (78, 7.8);
  \fill[gray!5] (-8, 4.6) rectangle (78, 5.8);
  \fill[gray!5] (-8, 2.6) rectangle (78, 3.8);

  % --- Vertical grid ---
  \foreach \x in {10,20,30,40,50,60} {
      \draw[gray!18] (\x, 1.5) -- (\x, 7.8);
  }
  \draw[gray!18] (0, 1.5) -- (0, 7.8);

  % --- X axis ---
  \draw[gray!50] (0, 1.5) -- (68, 1.5);
  \foreach \x in {0,10,20,30,40,50,60} {
      \node[font=\normalsize, below=2pt, text=gray!60!black] at (\x, 1.5) {\x};
  }
  \node[font=\normalsize, text=gray!50!black] at (33, 0.8) {Total training time (min, 10 epochs)};

  % --- Bar geometry ---
  \def\bh{0.18}
  \def\sep{0.24}

  % --- Legend (horizontal, above chart) ---
  \fill[red!28, draw=red!45, line width=0.4pt] (13, 8.3) rectangle (15, 8.6);
  \node[font=\normalsize, right=1pt] at (15.2, 8.45) {PyTorch};
  \fill[blue!22, draw=blue!38, line width=0.4pt] (27, 8.3) rectangle (29, 8.6);
  \node[font=\normalsize, right=1pt] at (29.2, 8.45) {JAX (training)};
  \fill[blue!38, draw=blue!50, line width=0.4pt, pattern=north east lines, pattern color=blue!25]
      (43, 8.3) rectangle (45, 8.6);
  \node[font=\normalsize, right=1pt] at (45.2, 8.45) {JAX (JIT compile)};

  % --- Speedup column header ---
  \node[font=\normalsize\bfseries, text=gray!55!black, anchor=center] at (73, 8.45) {Speedup};

  % ============================================================
  % NON-GNN MODELS (PyTorch + JAX) — seed 123
  % ============================================================

  % ====== NRMS (y=7.2) ======
  \def\yy{7.2}
  \node[font=\normalsize\bfseries, anchor=east] at (-0.8, \yy) {NRMS};
  \fill[red!28, draw=red!45, line width=0.4pt]
      (0, \yy+\sep-\bh) rectangle (29.16, \yy+\sep+\bh);
  \node[font=\normalsize, right=2pt, red!45!black] at (29.16, \yy+\sep) {29};
  \fill[blue!22, draw=blue!38, line width=0.4pt]
      (0, \yy-\sep-\bh) rectangle (12.83, \yy-\sep+\bh);
  \fill[blue!38, draw=blue!50, line width=0.4pt, pattern=north east lines, pattern color=blue!25]
      (12.83, \yy-\sep-\bh) rectangle (13.47, \yy-\sep+\bh);
  \node[font=\normalsize, right=2pt, blue!45!black] at (13.47, \yy-\sep) {13};
  \node[font=\normalsize\bfseries, text=green!45!black, anchor=center] at (73, \yy) {2.2$\times$};

  % ====== NAML (y=6.2) ======
  \def\yy{6.2}
  \node[font=\normalsize\bfseries, anchor=east] at (-0.8, \yy) {NAML};
  \fill[red!28, draw=red!45, line width=0.4pt]
      (0, \yy+\sep-\bh) rectangle (27.69, \yy+\sep+\bh);
  \node[font=\normalsize, right=2pt, red!45!black] at (27.69, \yy+\sep) {28};
  \fill[blue!22, draw=blue!38, line width=0.4pt]
      (0, \yy-\sep-\bh) rectangle (20.91, \yy-\sep+\bh);
  \fill[blue!38, draw=blue!50, line width=0.4pt, pattern=north east lines, pattern color=blue!25]
      (20.91, \yy-\sep-\bh) rectangle (22.59, \yy-\sep+\bh);
  \node[font=\normalsize, right=2pt, blue!45!black] at (22.59, \yy-\sep) {23};
  \node[font=\normalsize\bfseries, text=green!45!black, anchor=center] at (73, \yy) {1.2$\times$};

  % ====== LSTUR (y=5.2) ======
  \def\yy{5.2}
  \node[font=\normalsize\bfseries, anchor=east] at (-0.8, \yy) {LSTUR};
  \fill[red!28, draw=red!45, line width=0.4pt]
      (0, \yy+\sep-\bh) rectangle (13.70, \yy+\sep+\bh);
  \node[font=\normalsize, right=2pt, red!45!black] at (13.70, \yy+\sep) {14};
  \fill[blue!22, draw=blue!38, line width=0.4pt]
      (0, \yy-\sep-\bh) rectangle (11.39, \yy-\sep+\bh);
  \fill[blue!38, draw=blue!50, line width=0.4pt, pattern=north east lines, pattern color=blue!25]
      (11.39, \yy-\sep-\bh) rectangle (12.19, \yy-\sep+\bh);
  \node[font=\normalsize, right=2pt, blue!45!black] at (12.19, \yy-\sep) {12};
  \node[font=\normalsize\bfseries, text=green!45!black, anchor=center] at (73, \yy) {1.1$\times$};

  % ====== CAUM (y=4.2) ======
  \def\yy{4.2}
  \node[font=\normalsize\bfseries, anchor=east] at (-0.8, \yy) {CAUM};
  \fill[red!28, draw=red!45, line width=0.4pt]
      (0, \yy+\sep-\bh) rectangle (52.20, \yy+\sep+\bh);
  \node[font=\normalsize, right=2pt, red!45!black] at (52.20, \yy+\sep) {52};
  \fill[blue!22, draw=blue!38, line width=0.4pt]
      (0, \yy-\sep-\bh) rectangle (45.23, \yy-\sep+\bh);
  \fill[blue!38, draw=blue!50, line width=0.4pt, pattern=north east lines, pattern color=blue!25]
      (45.23, \yy-\sep-\bh) rectangle (46.57, \yy-\sep+\bh);
  \node[font=\normalsize, right=2pt, blue!45!black] at (46.57, \yy-\sep) {47};
  \node[font=\normalsize\bfseries, text=green!45!black, anchor=center] at (73, \yy) {1.1$\times$};

  % ====== PP-Rec (y=3.2) ======
  \def\yy{3.2}
  \node[font=\normalsize\bfseries, anchor=east] at (-0.8, \yy) {PP-Rec};
  \fill[red!28, draw=red!45, line width=0.4pt]
      (0, \yy+\sep-\bh) rectangle (63.50, \yy+\sep+\bh);
  \node[font=\normalsize, right=2pt, red!45!black] at (63.50, \yy+\sep) {64};
  \fill[blue!22, draw=blue!38, line width=0.4pt]
      (0, \yy-\sep-\bh) rectangle (27.51, \yy-\sep+\bh);
  \fill[blue!38, draw=blue!50, line width=0.4pt, pattern=north east lines, pattern color=blue!25]
      (27.51, \yy-\sep-\bh) rectangle (29.78, \yy-\sep+\bh);
  \node[font=\normalsize, right=2pt, blue!45!black] at (29.78, \yy-\sep) {30};
  \node[font=\normalsize\bfseries, text=green!45!black, anchor=center] at (73, \yy) {2.1$\times$};

  % ====== TCCM (y=2.2) ======
  \def\yy{2.2}
  \node[font=\normalsize\bfseries, anchor=east] at (-0.8, \yy) {TCCM};
  \fill[red!28, draw=red!45, line width=0.4pt]
      (0, \yy+\sep-\bh) rectangle (61.90, \yy+\sep+\bh);
  \node[font=\normalsize, right=2pt, red!45!black] at (61.90, \yy+\sep) {62};
  \fill[blue!22, draw=blue!38, line width=0.4pt]
      (0, \yy-\sep-\bh) rectangle (25.15, \yy-\sep+\bh);
  \fill[blue!38, draw=blue!50, line width=0.4pt, pattern=north east lines, pattern color=blue!25]
      (25.15, \yy-\sep-\bh) rectangle (27.56, \yy-\sep+\bh);
  \node[font=\normalsize, right=2pt, blue!45!black] at (27.56, \yy-\sep) {28};
  \node[font=\normalsize\bfseries, text=green!45!black, anchor=center] at (73, \yy) {2.2$\times$};

  \end{tikzpicture}
  }
  \caption{\small Total training time (minutes, 10 epochs) on MIND-small for non-GNN models (seed 123). Top bar is PyTorch, bottom bar is JAX split into training (solid) and one-time JIT compilation (hatched). JAX achieves 1.1--2.2$\times$ overall speedups.}
  \Description{Horizontal bar chart comparing total training time in minutes over 10 epochs for six non-GNN models on MIND-small. For each model a PyTorch bar is shown above a JAX bar, the latter split into training time and one-time JIT compilation. Values are NRMS 29 versus 13 minutes (2.2 times speedup), NAML 28 versus 23 (1.2 times), LSTUR 14 versus 12 (1.1 times), CAUM 52 versus 47 (1.1 times), PP-Rec 64 versus 30 (2.1 times), and TCCM 62 versus 28 (2.2 times).}
  \label{fig:training_time}
\end{figure}

\section{Conclusion}

We presented \textit{NewsReX}, an open-source multi-framework library for neural news recommendation that unifies distinct models across PyTorch and JAX/Flax behind a declarative YAML specification. The library provides shared, framework-agnostic infrastructure for graph construction, data processing, and evaluation, with only the forward-pass encoder layers being framework-specific. This design lets researchers leverage JAX's JIT speedups for attention-based models and PyTorch's graph ecosystem for GNN architectures, while sharing all other code. Pre-trained weights on Hugging Face allow building on existing results without redundant retraining. We reported multi-seed benchmarks and demonstrated multilingual generalization on a proprietary Japanese dataset from Nikkei News.

\subsection{Future Work}

Future work includes integration with pre-trained language models, expanding the model catalog to include more LLM-augmented architectures \cite{plm_rec, ip2}, more model implementations \cite{li-etal-2022-miner, cne_sue, lancer}, MIND-Large \cite{wu2020mind} and popular datasets like Adressa~\cite{adressa}. \textit{NewsReX} is actively maintained, and we will continue to add models, datasets, and training objectives as well as track upstream framework releases. Because implementations and dependencies evolve, metrics may shift between versions; the library version accompanying this paper can be tracked using the tag \texttt{v1.0.0-cikm26}.

\begin{acks}
This work was supported by JSPS KAKENHI Grant Number JP23K28908.
\end{acks}

\section*{GenAI Usage Disclosure}

Claude Code~\cite{claude-code} (Anthropic) was used to assist with verifying cross-framework implementation parity between PyTorch and JAX/Flax model implementations and to help monitor experiment runs conducted with \textit{NewsReX}. All generated outputs were reviewed and validated by the authors.

%%
%% The next two lines define the bibliography style to be used, and
%% the bibliography file.
\bibliographystyle{ACM-Reference-Format}
\bibliography{references}

\end{document}